\documentclass[aps,reprint, pre]{revtex4-1} 
\usepackage{caption,subcaption,amsmath,amssymb,amsfonts, graphics,graphicx,color,multirow,verbatim} %

\newcommand{\rmd}{\mathrm{d}}

\newcommand{\bK}{\mathbf{K}}
\newcommand{\bB}{\mathbf{B}}

\newcommand{\dxdt}{\dot{x}}
\newcommand{\ddxdt}{\ddot{x}}

\newcommand{\bkappa}{\boldsymbol{\kappa}}
\newcommand{\bb}{\mathbf{b}}
\newcommand{\bu}{\mathbf{u}}
\newcommand{\tzeta}{\tilde{\zeta}}
\newcommand{\tlambda}{\tilde{\lambda}}
\newcommand{\bogamma}{\boldsymbol{\gamma}}

\newcommand{\at}{\makeatletter @\makeatother}

\begin{document}
\title{Fluctuating, Lorentz-force-like coupling of Langevin equations and heat flux rectification}
\author{B. Sabass}
\affiliation{Institute of Complex Systems 2, Forschungszentrum J{\"u}lich, J{\"u}lich, Germany}
\email{B.Sabass\at fz-juelich.de}
\begin{abstract}
In a description of physical systems with Langevin equations, interacting degrees of freedom are usually coupled through symmetric parameter matrices. This coupling symmetry is a consequence of time-reversal symmetry of the involved conservative forces. If coupling parameters fluctuate randomly, the resulting noise is called multiplicative. For example, mechanical oscillators can be coupled through a fluctuating, symmetric matrix of spring ``constants''. Such systems exhibit well-studied instabilities. In this note, we study the complementary case of antisymmetric, time-reversal symmetry breaking coupling that can be realized with Lorentz forces or various gyrators. We consider the case that these antisymmetric couplings fluctuate. This type of multiplicative noise does not lead to instabilities in the stationary state but renormalizes the effective non-equilibrium friction. Fluctuating Lorentz-force-like couplings also allow to control and rectify heat transfer. A noteworthy property of this mechanism of producing asymmetric heat flux is that the controlling couplings do not exchange energy with the system.
\end{abstract}
\maketitle
\section{Introduction}
Continuous stochastic processes can be modeled through differential equations with added noise processes. If a noise process appears in a product with a function of the system variables, noise is referred to as multiplicative. The study of multiplicative noise has a long history since it can cause rather dramatic phenomena~\cite{van1976stochastic}. For example, even arbitrarily weak stochastic fluctuations of the eigenfrequency in harmonic oscillator models lead to instabilities in higher moments of system variables~\cite{bourret1971energetic, lindenberg1984finite}. Similarly, fluctuating friction parameters can prohibit stable stationary solutions~\cite{luczka2000diffusion, mallick2002anomalous, gitterman2005classical,mendez2011instabilities, gitterman2013mass}. Such ``energetic instabilities''\cite{bourret1971energetic} occur since forces resulting from fluctuating potentials or friction parameters pump energy in and out of the system. So far, multiplicative noise processes have been studied either for one-dimensional systems or for forces that couple different degrees of freedom symmetrically. In this note, we consider couplings that are antisymmetric under time reversal and thus lead to antisymmetric coupling matrices. Stochastic changes of these ``Lorentz-Force-like'' couplings produces multiplicative noise.

Lorentz-forces can not perform work or change the internal energy since they always act normal to velocities. Therefore, fluctuating Lorentz-force-like couplings yield a special type of multiplicative noise that is energetically neutral. Below, we derive generic differential equations governing the first and second moments of linear systems with fluctuating Lorentz-force-like couplings. It is shown that this type of multiplicative noise does not lead to instabilities but increases the effective friction that damps the first moment when external forces are applied. Fluctuations in Lorentz-force-like couplings do not affect equilibrium correlations between different degrees of freedom but modify the non-equilibrium correlations. 

Next, the energetics of our systems are studied within the framework of stochastic thermodynamics~\cite{sekimoto1998langevin,SeifertReview}. On the level of Langevin equations, the first law of thermodynamics naturally leads to a definition of heat. Assuming that different degrees of freedom are exposed to separate thermal environments with different temperature, we can calculate heat transfer through the system. This heat transfer can be controlled by fluctuating Lorentz-force-like couplings because they modify the non-equilibrium correlations. As an example we analyze heat transfer in a two-component system. Both components are in contact with their own heat bath, which fixes the additive noise strengths to different values. Random motion of one component is transmitted via Lorentz-force-like coupling to the other, thus, heat is transmitted. Finally, the model is augmented by the assumption that the fluctuation strength of the multiplicative noise in the coupling is also determined by one of the baths. For this system, heat transfer is no longer symmetric under reversal of the temperature difference and the system acts as a rectifier for heat. This mechanism of rectifying heat transfer is notable for the fact that the Lorentz-force-like couplings producing the heat transfer asymmetry do not exchange any energy with the system.
\section{Fluctuating, antisymmetric coupling of Langevin equations}
\subsection{The Langevin equations}
In the following, all quantities are assumed to be non-dimensional and the Boltzmann constant $k_{\rm b}$ is set to unity. The Einstein summation convention is not employed. We study a system of coupled, time dependent, real variables $x_j(t)$ that could, e.g., represent the positions of microscopic particles or the charge of electric oscillators. In such systems, the time derivatives $\dxdt_j$, i.e., the velocities or currents, can be coupled through Lorentz forces or Coriolis forces that break time-reversal symmetry. A general form of the Langevin equations governing the $x_j$ is
\begin{align}
\ddxdt_j = -\sum_l\left[\kappa_{jl}\,x_l + (b_{jl}+\tilde{\zeta}_{jl})\,\dxdt_l + \gamma_{jl}\,\dxdt_l\right]+\xi_j + f_j.
\label{eq_network}
\end{align}
The symmetric matrix $\bkappa=\bkappa^{T}$ represents, e.g., spring constants in a mechanical system or capacitance in an electric network. $\bkappa$ is to be positive definite for stability~\cite{strogatz2006nonlinear}. We thereby also exclude the marginally stable case where one eigenvalue of $\bkappa$ is zero. The antisymmetric matrix $\bb = -\bb^{T}$ represents Lorentz-force-like couplings which are, e.g., realizable through a magnetic field. Fluctuations in the antisymmetric couplings are modeled by the noise matrix $\tilde{\boldsymbol{\zeta}} = -\tilde{\boldsymbol{\zeta}}^{T}$. The multiplicative noise $\sim\tilde{\zeta}_{jl}\dxdt_l$ is interpreted in the Stratonovich sense. Finally, we also have a positive definite, symmetric ``friction matrix'' $\bogamma=\bogamma^{T}$. The two last quantities on the right side of Eq.~(\ref{eq_network}) are the thermal noise $\xi_j$ and a time-dependent force $f_j$. 

The statistical average is written as $\langle\ldots \rangle$. Both types of fluctuations have zero average as $\langle\xi_{j}(t)\rangle=0$ and $\langle\tzeta_{jl}(t)\rangle=0$. Different types of fluctuations are to be independent, thus, $\langle\tzeta_{jl}(t)\,\xi_{k}(t')\rangle=0$. For many physical systems, the fluctuation autocorrelations decay exponentially. Such Ornstein-Uhlenbeck-type correlations with inverse relaxation times $\lambda$ and $\tlambda$ read for $t \geq 0$ 
\begin{subequations}
\begin{align}
\langle\xi_{j}(0)\xi_{j'}(t)\rangle &= \frac{\lambda}{2}\, e^{-\lambda t} K_{j,j'},\label{eq_c1}\\
\langle\tzeta_{jl}(0)\tzeta_{j'l'}(t)\rangle &= \frac{\tlambda}{2}\,e^{-\tlambda t} B_{jl}(\delta_{j,j'}\delta_{l,l'}-\delta_{j,l'}\delta_{l,j'})\label{eq_c2},
\end{align}
\end{subequations}
The symmetric, positive matrix $\bK$ in Eq.~(\ref{eq_c1}) determines the strength of the additive noise. Analogously, $\bB$ in Eq.~(\ref{eq_c2}) determines the strength of the multiplicative noise. This matrix is symmetric $\bB=\bB^{T}$, has only positive entries $B_{ij}\geq 0$, and zeros on the diagonal $B_{ii} =0$. 
For simplicity, we will focus in the following on the white noise limit of Eqns.~(\ref{eq_c1},\ref{eq_c2}) where
\begin{align}
\lambda \rightarrow \infty,& & \tlambda\rightarrow \infty.\label{eq_lim_lambdas}
\end{align}
In this limit we have $\lim_{\lambda\rightarrow\infty}(\lambda\, e^{-\lambda t}/2) = \lim_{\tlambda\rightarrow\infty}(\tlambda\, e^{-\tlambda t}/2) =\delta(t)$. For Gaussian noise, cumulants with order higher than two vanish and we can express noise correlations through products of pairwise correlations.
\subsection{General solution for the first moment}
Taking the average $\langle\ldots\rangle$ of Eq.~(\ref{eq_network}) yields 
\begin{align}
\langle \ddxdt_j \rangle &= - \sum_l \left[\kappa_{jl} \langle x_l \rangle +(b_{jl}+\gamma_{jl})\langle \dxdt_l\rangle \right]-\sum_l \langle\tzeta_{jl} \dxdt_l\rangle + f_j,
\label{eq_mean_with_noise_raw}
\end{align}
which leaves us with the problem of calculating the expectation value of correlations with the system variables of form $\langle\tzeta_{jl} \dxdt_l\rangle$. For the case of exponentially decaying, Gaussian noise-noise correlations, solutions exist in the form of a systematic expansion for short correlation times~\cite{bourret1971energetic, van1974cumulant, shapiro1978formulae}. Here, we consider the white noise limit $\tlambda\rightarrow \infty$ and using noise splitting formulas detailed in Ref.~\cite{shapiro1978formulae} we obtain $\langle\tzeta_{jl} \dxdt_l \rangle= B_{jl}\langle\dxdt_j\rangle/2$ (see Appendix). Thus, the first moments obey
\begin{align}
\langle \ddxdt_j \rangle &= - \sum_l \left[\kappa_{jl} \langle x_l \rangle +(b_{jl}+\gamma_{jl})\langle \dxdt_l\rangle \right]-\sum_m \frac{B_{jm}}{2}\langle \dxdt_j\rangle + f_j.
\label{eq_mean_with_noise}
\end{align}
Here, the term $-\sum_m B_{jm}\langle \dxdt_j\rangle/2$ increases the ``friction'' on the average trajectories because $\bB$ is positive~\cite{sabass2015network}. The renormalization of the friction constants can possibly be interpreted as a geometric effect since a Lorentz-force produces ``curved trajectories''. Positivity of the effective friction in Eq.~(\ref{eq_mean_with_noise}) is a consequence of the antisymmetry of the Lorentz-force-like couplings $\bb=-\bb^{T}$, which appears in the Kronecker-delta expression in Eq.~(\ref{eq_c2}) as antisymmetry under index exchange $j \leftrightarrow l$. Note that fluctuations in the friction parameters $\gamma$ produce the opposite effect, namely a reduced effective friction ~\cite{gitterman2005classical}; which can lead to unstable stationary solutions when the effective friction becomes negative.
\subsection{General solution for the second moment}
The equations governing the second moments result from multiplying Eq.~(\ref{eq_network}) with derivatives of $x_k$ and subsequent averaging. A lengthy calculation yields
\begin{subequations}
\begin{align}
&\frac{\rmd}{\rmd t}\langle x_m x_k\rangle = \langle \dxdt_k x_m\rangle + \langle x_k \dxdt_m\rangle, \label{eq_fluct_b_secondmoment1}
\\
\begin{split}
&
\frac{\rmd}{\rmd t}\langle x_m \dxdt_k \rangle = \langle \dxdt_m \dxdt_k \rangle - \sum_j \frac{B_{kj}}{2}\langle x_m \dxdt_k \rangle\\
&- \sum_j(\kappa_{kj}\langle x_m x_j \rangle+ [b_{kj} + \gamma_{kj}]\langle x_m \dxdt_j \rangle)+\langle x_m \rangle f_k,\label{eq_fluct_b_secondmoment2}
\end{split}
\\
\begin{split}
&\frac{\rmd}{\rmd t}\langle \dxdt_m \dxdt_k \rangle = K_{mk}+\sum_j \delta_{mk} B_{kj}\langle \dxdt_j^2 \rangle\\
&-\sum_j ( \frac{B_{kj}}{2} + \frac{B_{mj}}{2})\langle \dxdt_m \dxdt_k\rangle - B_{km}\langle\dxdt_m \dxdt_k\rangle\\
&- \sum_j\left(\kappa_{kj}\langle \dxdt_m x_j \rangle +[b_{kj}+\gamma_{kj}]\langle \dxdt_m \dxdt_j\rangle\right)+ \langle \dxdt_m\rangle f_k\\
&- \sum_i\left(\kappa_{mi}\langle \dxdt_k x_i \rangle +[b_{mi}+\gamma_{mi}]\langle \dxdt_k \dxdt_i\rangle\right)+\langle \dxdt_k\rangle f_m.\label{eq_fluct_b_secondmoment3}
\end{split}
\end{align}
\end{subequations}
The equations~(\ref{eq_mean_with_noise},\ref{eq_fluct_b_secondmoment1}-\ref{eq_fluct_b_secondmoment3}) for the first and second moments form a closed system that can readily be solved. The following provides an example involving the calculation of heat exchange.
\section{Heat exchange}
From now on the friction matrix in Eq.~(\ref{eq_network}) is assumed to be diagonal $\gamma_{jl}=\delta_{jl}\gamma_{j}$. Furthermore, we assume that the noise processes $\xi_{j}$ in the Langevin equations result from thermal equilibrium fluctuations of large baths that surround the individual degrees of freedom $x_j$. The coupling between the $x_j$ and the baths should not depend on the system state and the bath fluctuations are independent of the system. Each $x_j$ is connected to its own bath with temperature $T_j$. Therefore, correlations of the noise variables obey~\cite{Kubo1992statistical,sekimoto1998langevin}
\begin{align}
K_{jl}=\delta_{jl}\,2 \, \gamma_{j} T_j.
\label{eq_fdt1}
\end{align} 
\subsection{Definition of heat}
The Langevin dynamics can be endowed straight-forwardly with a thermodynamical interpretation as follows. We multiply Eq.~(\ref{eq_network}) by $\dxdt_j$ and subsequently sum over $j$. The antisymmetric coupling matrices do not appear in this balance equation since $\sum_{jl}\dxdt_j(b_{jl}+\tilde{\zeta}_{jl})\,\dxdt_l=0$. Therefore, these coupling forces do not affect the energetics. After averaging, we obtain the following energy balance
\begin{align}
\begin{split}
\sum_{j l} \frac{\rmd}{\rmd t}[\delta_{jl}\langle \frac{\dxdt^2_j}{2} \rangle +\langle \frac{x_j \kappa_{jl} x_l}{2} \rangle] = 
\sum_j \langle f_j \dxdt_j \rangle -(\gamma_{j}\langle \dxdt^2_j \rangle - \frac{K_{jj}}{2}),
\end{split}
\label{eq_first_law}
\end{align}
where the expression on the left hand side is the change of internal energy. The first term on the right hand side is the work done by the forces $\mathbf{f}$. The second term on the right hand side is the average momentum exchange with the temperature baths. Accordingly, the heat exchange of each element $j$ with its thermal environment is defined as~\cite{sekimoto1998langevin, harada2006energy}
\begin{align}
\dot{Q}_{j} \equiv \gamma_{j}\langle\dxdt_j^2\rangle - \frac{1}{2}K_{jj}.
\label{eq_heat}
\end{align}
Note that this definition leads to heat fluxes that are linear combinations of the temperatures $\dot{Q}_{j}=\alpha_1 \,T_1 + \alpha_2\,T_{2} + \ldots$ where the coefficients satisfy $\sum_{k}\alpha_k =0$. The latter constraint reduces the number of variables by one, such that $\dot{Q}_{j}$ can always be written as function of the temperature differences. 

To describe an equilibrium situation we set $f_j=0$ and require that all fluctuations are determined by a single temperature $T_{\rm eq}$ such that the noise correlations for all $j$, $l$ are given by 
\begin{align}
K^{\rm eq}_{jl} = \delta_{jl}\, 2\,\gamma_{j} T_{\rm eq}.
\end{align}
For long times, stationary correlation functions result from Eqns.~(\ref{eq_fluct_b_secondmoment2}, \ref{eq_fluct_b_secondmoment3}) as $\langle \dxdt_k x_m\rangle_{\mathrm{eq}} = \langle x_k \dxdt_m\rangle_{\mathrm{eq}} =0$ and $\langle \dxdt_m \dxdt_k \rangle_{\mathrm{eq}} = \delta_{mk}T_{\rm eq}$. Thus, the multiplicative noise strength $\bB$ becomes irrelevant in equilibrium. Although fluctuations in the antisymmetric coupling matrices do not change the internal energy or produce work, they do affect the transfer of energy between different degrees of freedom in non-equilibrium.
\begin{figure}[htb]
        \centering
        \begin{subfigure}[htb]{0.43\textwidth}
                \includegraphics[width=\textwidth]{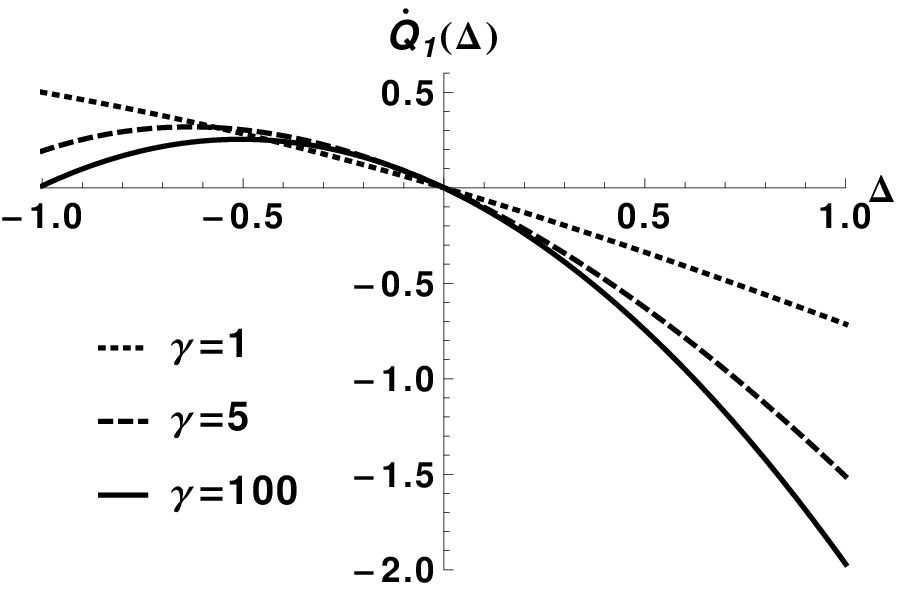}
        \end{subfigure}%
				\\ 
				\vspace{0.7cm}
        \begin{subfigure}[htb]{0.43\textwidth}
                \includegraphics[width=\textwidth]{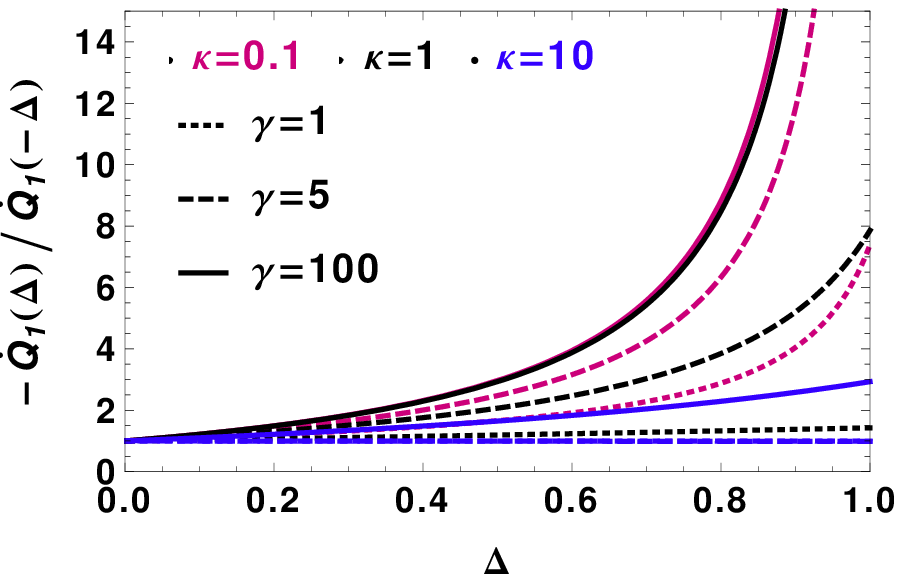}
        \end{subfigure}
				\caption{Top: Heat transfer vs. temperature difference $\Delta = (T_1-T_2)/2$ for $\kappa=1$. Bottom: Asymmetry of heat transfer.
				Heat transfer in Eq.~(\ref{eq_example_two_oszis}) can become asymmetric when the multiplicative noise strength $B_{12}$ depends on temperature through Eq.~(\ref{eq_temperature_dependent_B12}). Here $T_1 = (1+\Delta)$, $T_2 = (1-\Delta)$, and $\nu=1$.
}
\label{fig_adg}
\end{figure}
\subsection{A toy model for heat flux control}
We next consider an example for how the multiplicative noise $\tzeta$ can allow to control heat transfer. The general Langevin equation~(\ref{eq_network}) is specialized to the case of two elements. Furthermore, the system is simplified by assuming a stationary state with $f_j=0$ and by assuming that the magnetic coupling is on average zero ($\bb =\mathbf{0}$). The governing equations are
\begin{align}
\begin{split}
\begin{pmatrix}
\ddxdt_1\\
\ddxdt_2\\
\end{pmatrix}=2
\begin{pmatrix}
-2\kappa & \kappa\\
\kappa & -2\kappa
\end{pmatrix}
\begin{pmatrix}
x_1\\
x_2
\end{pmatrix}
-\begin{pmatrix}
\gamma  & \tzeta \\
-\tzeta & \gamma 
\end{pmatrix}
\begin{pmatrix}
\dxdt_1\\
\dxdt_2\\
\end{pmatrix}
+\begin{pmatrix}
\xi_1\\
\xi_2\\
\end{pmatrix}.
\label{eq_example_two_oszis}
\end{split}
\end{align} 
The two oscillators are to be connected with different heat baths at temperatures $T_1$ and $T_2$. Thus, the strength of the additive noise $\xi_{\{1,2\}}$ is determined by 
\begin{align}
K_{11} = 2\gamma T_1, & & K_{22} = 2\gamma T_2, & & K_{12} = K_{21} = 0.
\end{align}
The stationary heat exchange can now be calculated straightforwardly from Eqns.~(\ref{eq_fluct_b_secondmoment3},\ref{eq_heat},\ref{eq_fdt1}). The result is
\begin{align}
\dot{Q}_1 = \frac{\gamma (2 \kappa + B_{12} (2 \gamma + B_{12})) (T_2 - T_1)}{4 \kappa + 2 (\gamma + B_{12}) (2 \gamma + B_{12})}, & & \dot{Q}_2 =-\dot{Q}_1.
\label{eq_heat_white_mult_noise}
\end{align}
Clearly, heat is a non-linear function of the multiplicative noise strenght $B_{12}$. However, the multiplicative noise can not change the direction of heat transfer in Eq.~(\ref{eq_heat_white_mult_noise}) since per definition $B_{12}\geq0$. Thus, spontaneous currents from the colder to the hotter heat bath can not occur and thermodynamic consistency is retained. This result is a consequence of the usage of antisymmetric couplings as fluctuating quantity since no energy is injected or removed during fluctuations. 

The non-monotonous dependence of $\dot{Q}$ on the coupling constants in Eq.~(\ref{eq_heat_white_mult_noise}) allows to
control heat transfer through the strength of the multiplicative noise. The two extreme limits of vanishing and very strong multiplicative noise yield a heat transfer of $\dot{Q}_1|_{B_{12}=0} = \gamma \kappa (T_2 - T_1)/(2\gamma^2 + 2\kappa)$ and $\dot{Q}_1|_{B_{12}\rightarrow\infty} \approx \gamma (T_2 - T_1)/2$. In between these limits, a minimum occurs at the fluctuation strength $B_{12}=\sqrt{2 \kappa} - 2 \gamma \geq 0$ with a heat transfer of
\begin{equation}
\dot{Q}_1^{\mathrm{min}} = \frac{\gamma ( \sqrt{2 \kappa}/\gamma - 1)}{2 \sqrt{2 \kappa}/\gamma - 1}(T_2 - T_1)\geq\frac{\gamma}{3}(T_2 - T_1).
\end{equation}
For large friction constants, when $\gamma \geq \sqrt{\kappa/2}$, the minimum heat transfer occurs at $B_{12}=0$ and $\dot{Q}_1$ increases monotonously with multiplicative noise intensity.
\subsection{Rectification of heat exchange}
Instead of fixing $\bB$ to some arbitrary value, we now assume that the fluctuations $\tzeta$ in Eq.~(\ref{eq_example_two_oszis}) are governed by the temperature $T_1$ of one of the heat baths. In this case, $B_{12}$ is proportional to the temperature as
\begin{align}
B_{12} = B_{21} = \nu\,T_1, & & B_{11} = B_{22} = 0,\label{eq_temperature_dependent_B12}
\end{align}
where $\nu$ is a constant. With this definition of $\bB$, Eq.~(\ref{eq_heat_white_mult_noise}) yields a nonlinear dependence of $\dot{Q}_{1,2}$ on $T_1$. Moreover, the magnitude of heat exchange depends asymmetrically on the direction of heat transfer $T_1 \leftrightarrow T_2$. Fig.~\ref{fig_adg}a) shows plots of Eq.~(\ref{eq_heat_white_mult_noise}) for symmetric temperature difference $T_{1,2} = 1 \pm \Delta$. As demonstrated in the figure, heat transfer becomes a quadratic function $\dot{Q}_1\approx -(\Delta + \Delta^2) \nu$ when $\gamma\gg\nu$ and also $\gamma\gg\kappa$. Then, the magnitude of heat transfer in the direction $T_1 \rightarrow T_2$ ($\Delta >0$) is stronger than in the reverse direction. 

To quantify the asymmetry of heat transfer we consider the quantity $\dot{Q}_1(\Delta)/\dot{Q}_1(-\Delta)$ in Fig.~\ref{fig_adg}b). The asymmetry becomes large when $\Delta\simeq 1$, i.e., when the temperature difference is comparable to the mean temperature $(T_1+T_2)/2$. The plot also demonstrates that increasing the coupling constant $\kappa$ generally leads to a less pronounced asymmetry in the heat transfer.
\section{Concluding remarks}
Lorentz-force-like couplings can be physically realized in different ways. Devices that couple fluxes in a non-reciprocal way are known in electrical engineering as gyrators~\cite{tellegen1948gyrator} and early designs were based on a rectangular Hall element with  separate ports at all four sides~\cite{wick1954solution}. Recent developments include gyrators based on magnetoelectric materials~\cite{zhai2006quasi} and Hall effect gyrators with significantly reduced electrical resistance \cite{viola2014hall,bosco2017nonreciprocal}. It is also possible to build non-reciprocal Microwave wave guides by exploiting the Faraday effect~\cite{hogan1952ferromagnetic}. While these devices rely on time-reversal symmetry-breaking properties of magnetic fields, one can in principle imagine replacing the Lorentz forces by Coriolis forces in a rotating inertial frame. Fluctuations in the resulting antisymmetric couplings could then be caused, e.g., by noise in the magnetic field or in the angular velocity. While a detailed study of the resulting phenomena would require solving Maxwell's equations or mechanical force balance equations, we focus in this note on generic second order stochastic differential equations with fluctuating, antisymmetric coupling of the velocity variables. For these systems, we derive general equations governing the first and second moments under the assumption of Gaussian white noise. It is demonstrated that the new type of multiplicative noise only affects the out-of-equilibrium correlations and does not lead to energetic instabilities. As an application of our formulas we discuss heat transport through a system with fluctuating Lorentz-force like couplings.

Any heat transport process must satisfy the second law of thermodynamics, i.e., heat does not flow spontaneously from a cooler reservoir to a hotter reservoir~\cite{clausius1854veranderte}. Fourier's law of heat conduction is linear in temperature differences and therefore satisfies the requirement naturally. However, non-linear and asymmetric heat conduction laws are also possible. Using our framework for multiplicative noise in Lorentz-force-like couplings, we study heat transfer between two reservoirs. Noise processes in the Langevin equations can be interpreted as thermal equilibrium fluctuations in a temperature bath. This assumption leads to a natural microscopic identification of heat and work in the stochastic system, whereby the nonequilibrium heat exchanged between the bath and the system is related to the velocity autocorrelations. Consequently, heat flow can be controlled through fluctuations of the Lorentz-force-like coupling. This way of controlling heat flow automatically conserves the energy balance due to energetic neutrality of antisymmetric coupling matrices. Therefore, such systems can be studied consistently without explicitly modeling the origin of the multiplicative noise by additional equations, which presents an advantage for theoretical work.

Concepts for rectification of heat flow have received considerable scientific attention during recent years. In particular, studies of low-dimensional nanoscale-systems yielded various principles that allow to control heat flux and produce asymmetry under exchange of the heat flow direction~\cite{terraneo2002controlling, li2004thermal, hu2006asymmetric, dhar2008heat, li2012colloquium, cahill2014nanoscale, li2006negative,wang2007thermal, chung2008thermal, segal2008single, segal2008nonlinear,joulain2016quantum}. Studying different instances of heat flux rectification is not only important for an understanding of general principles, but may also have immediate applications, for example in nanotechnology.
\begin{acknowledgments}
This article was written during a postdoctoral stay at Princeton University. The author cordially thanks H.A.~Stone and Z.~Gitai for support and acknowledges a postdoctoral fellowship from the German Academic Exchange Service (DAAD).
\end{acknowledgments}
\section*{Appendix}
In the following, we demonstrate the use of standard procedures to calculate the noise correlators employed above.
For a system with $N$ space coordinates we define $\bu \equiv \{x_1\ldots x_N,\dxdt_1\ldots \dxdt_N\}$ to write Eq.~(\ref{eq_network}) as a system of first order differential equations
\begin{align}
\frac{\rmd u_j}{\rmd t} = \sum_l\left[A_{jl}\,u_l + Z_{jl}\,u_l\right] +\Xi_j + F_j.
\label{eq_network_compact}
\end{align}
Here, the constant parameters $\bkappa$, $\mathbf{b}$, and $\bogamma$ in Eq.~(\ref{eq_network}) are absorbed in $A_{jl}$. The additive noise is $\Xi_j =0$ for $j \leq N$ and $\Xi_j =\xi_{j-N}$ for $j>N$. External forces acting on the system are contained in $F_j$. The matrix $Z_{jl}$ contains the multiplicative noise and its entries are
\begin{align*}
&Z_{jl} = -\tzeta_{(j-N),(l-N)} \,\,\, & \mathrm{for}\,\,j>N\,\,\mathrm{and}\,\, l>N\\
&Z_{jl} = 0 \,\,\, & \mathrm{otherwise.}
\end{align*}
Next, we consider an arbitrary function $h(t)$ that depends on the zero-mean stationary Gaussian noises $Z_{jl}$. We seek to calculate equal time correlations of the form $\langle Z_{..}(t) h(t)\rangle$. Following Ref.~\cite{shapiro1978formulae}, we expand $h(t)$ in time-ordered products of the noise variables through a functional Taylor series. As a slight generalization of results given in Ref.~\cite{shapiro1978formulae} we find 
\begin{subequations}
\begin{align}
&\frac{\rmd}{\rmd t}\langle Z_{jl} h\rangle=- \tilde{\lambda} \langle Z_{jl} h\rangle +\langle \tzeta_{jl} \frac{\rmd h}{\rmd t}\rangle,
\label{eq_shapiro}\\
\begin{split}
&\frac{\rmd}{\rmd t}\langle Z_{m n} Z_{jl} h\rangle=- 2\tilde{\lambda}\langle Z_{m n} Z_{jl} h\rangle\\ 
&+\langle Z_{m n} Z_{jl} \frac{\rmd h}{\rmd t}\rangle+2\tilde{\lambda}\langle Z_{m n} Z_{jl}\rangle \langle h\rangle,
\end{split}
\label{eq_shapiro2}\\
\begin{split}
&\frac{\rmd}{\rmd t}\langle Z_{rs} Z_{m n} Z_{jl} h\rangle=- 3\tilde{\lambda}\langle Z_{rs} Z_{mn} Z_{jl} h\rangle \\
&+ \langle Z_{rs} Z_{mn} Z_{jl} \frac{\rmd h}{\rmd t}\rangle + 2\tilde{\lambda}[ \langle Z_{r s} Z_{m n}\rangle \langle Z_{j l} h\rangle \\
&+ \langle Z_{rs} Z_{jl}\rangle \langle Z_{n m} h \rangle + \langle Z_{m n} Z_{jl}\rangle \langle Z_{r s} h\rangle],
\end{split}
\label{eq_shapiro3}
\end{align}
\end{subequations}
These formulas hold for both of our noise sources with exchanged variables $Z_{kl} \rightarrow \Xi_j$, $\tilde{\lambda} \rightarrow \lambda$. We can now set $h = u_j$ in Eqns.~(\ref{eq_shapiro}-\ref{eq_shapiro3}) and use the Langevin equation (\ref{eq_network_compact})
to obtain a hierarchy of equations where every correlation is connected to correlations of the next higher order in $Z_{..}$. Integration of Eqns.~(\ref{eq_shapiro},\ref{eq_shapiro2}) yields
\begin{subequations}
\begin{align}
\langle Z_{jl} u_k\rangle =& \int_0^t e^{-\tlambda (t -t')} \langle Z_{jl} \frac{\rmd u_k}{\rmd t} \rangle|_{t'}\, \rmd t',\\
\begin{split}
\langle Z_{jl} Z_{mn} u_k\rangle =& \int_0^{t'}  e^{-2\tlambda (t' -t'')} [2 \tlambda \langle Z_{jl} Z_{mn} \rangle \langle u_k \rangle\\
& +\langle Z_{jl} Z_{mn} \frac{\rmd u_k}{\rmd t} \rangle]_{t''}\rmd t'',
\end{split}
\end{align}
\end{subequations}
where we assumed that the contribution of initial values of the correlations vanishes. Together, these equations yield
\begin{align}
\begin{split}
&\langle Z_{jl} u_k\rangle = \sum_m \int_0^t e^{-\tlambda (t -t')} [ \langle Z_{jl} (A_{km} u_m + \Xi_k + F_k)\rangle|_{t'} + \\
&\int_0^{t'} e^{-2\tlambda (t' -t'')}(2 \tlambda \langle Z_{jl} Z_{km} \rangle \langle u_m \rangle + \langle Z_{jl} Z_{km} \frac{\rmd u_m}{\rmd t} \rangle)_{t''}\rmd t'' ] \, \rmd t',
\end{split}
\label{eq_dasisteinspass}
\end{align}
We next consider the limit of very short noise correlations $\tlambda \rightarrow \infty$ where of course $t>0$. Assuming that the sought-for correlations are finite, the terms in Eq.~(\ref{eq_dasisteinspass}) in the first line on the right side yield a contribution of vanishing measure since the factor $e^{-\tlambda (t -t')}$ is zero for $(t -t')>0$ and only finite for the point $(t -t')=0$. For the first summand on the second line of Eq.~(\ref{eq_dasisteinspass}), we can
employ the correlation relations (\ref{eq_c2}), yielding a contribution that can be at most $\sim \tlambda^2$. This term survives the limit of $\tlambda \rightarrow \infty$ since then $\tlambda e^{-\tlambda (t -t')} \rightarrow 2\delta(t-t')$.\\
To evaluate the last term in Eq.~(\ref{eq_dasisteinspass}) we could again replace $\frac{\rmd u_j}{\rmd t}$ by Eq.~(\ref{eq_network_compact}) and use the integral of Eq.~(\ref{eq_shapiro3}). However, the last three summands on the right hand side of Eq.~(\ref{eq_shapiro3}) are at most $\sim\tlambda^2$ and are therefore suppressed by the exponential integral factors in the limit of $\tlambda \rightarrow \infty$. The second term on the right side of Eq.~(\ref{eq_shapiro3}) can only yield a non-zero, finite contribution in the case that $\langle Z_{..}Z_{..}Z_{..}Z_{..} u_i\rangle \sim \lambda^3$. However, this case is rejected on physical grounds since for Gaussian noise $\langle Z_{..}^4\rangle$ is at most $\sim \tlambda^2$ and $u_i$ varies on a much longer timescale than the noise variable. Thus, the only non-vanishing contribution to the integral in Eq.~(\ref{eq_dasisteinspass}) comes from the first summand in the second line. The result is
\begin{align}
\begin{split}
&\langle Z_{jl} u_k\rangle = \\
&\sum_m \int_0^te^{-\tlambda (t -t')}\int_0^{t'} e^{-2\tlambda (t' -t'')} 2 \tlambda \langle Z_{jl} Z_{km} \rangle \langle u_m \rangle|_{t''} \rmd t'' \, \rmd t'.
\end{split}
\label{eq_dasisteinspass_final}
\end{align}
Next, we revert back to our original variables. Since the matrix elements $\{Z_{..} \}$ contain multiplicative noise components $\{-\tzeta_{..} \}$, we employ Eq.~(\ref{eq_c2}) and take the limit of large $\tlambda$ to obtain
\begin{subequations}
\begin{align}
\langle \tzeta_{jl} x_k\rangle &=0,\\
\langle \tzeta_{jl} \dxdt_k\rangle &= \frac{B_{jl}}{2}(\delta_{lk}-\delta_{jk})\langle\dxdt_k\rangle .
\end{align}
\end{subequations}
These are the correlations that were employed for derivation of Eq.~(\ref{eq_mean_with_noise}).
Through an analogous calculation for the additive noise $\boldsymbol{\xi}$ we obtain by simply exchanging the
noise variables in above derivation
\begin{subequations}
\begin{align}
\langle \xi_{j} x_k\rangle &=0,\\
\langle \xi_{j} \dxdt_k\rangle &= \frac{K_{jk}}{2}.
\end{align}
\end{subequations}

For calculation of the second moments of the system variables we need the correlation between noise and two system variables $\langle Z_{jl} u_k u_i \rangle$. This expression can be evaluated by setting $h=u_k u_i$ in Eqns.~(\ref{eq_shapiro},\ref{eq_shapiro2}) and by then following through with the same procedure as above. The final result, in original variables after evaluation of the Kronecker-delta expressions Eq.~(\ref{eq_c2}), reads
\begin{subequations}
\begin{align}
\langle\tzeta_{jl} x_k x_i\rangle =& 0,\\
\langle\tzeta_{jl} \dxdt_k x_i\rangle =& -\frac{B_{jl}}{2} [\delta_{jk} \langle \dxdt_l x_i\rangle - \delta_{lk} \langle \dxdt_j x_i\rangle],\\
\begin{split}
\langle\tzeta_{jl} \dxdt_k \dxdt_i\rangle =& -\frac{B_{jl}}{2} [\delta_{ji} \langle \dxdt_k \dxdt_l\rangle - \delta_{li} \langle \dxdt_k 
\dxdt_j\rangle]\\
&-\frac{B_{jl}}{2} [\delta_{jk} \langle \dxdt_l \dxdt_i\rangle - \delta_{lk} \langle \dxdt_j \dxdt_i\rangle].
\end{split}
\end{align}
\end{subequations}

%
\end{document}